\documentclass[twocolumn,amsmath,amssymb,nofootinbib]{revtex4}
\usepackage{graphicx}
\usepackage{dcolumn}
\usepackage{color}
\usepackage{scrextend}
\usepackage{subfig}
\usepackage{bm}
\usepackage{amsmath,amssymb,graphicx}
\usepackage{epsfig}
\usepackage{enumerate}
\usepackage{array}
\usepackage{multirow}
\usepackage{ulem}

\begin{document}
\title
{Brownian-motion approach to statistical mechanics:\\ Langevin equations, fluctuations, and timescales}
\author{Sushanta Dattagupta$^{1}$\footnote{sushantad@gmail.com} and Aritra Ghosh$^2$\footnote{ag34@iitbbs.ac.in}}
\affiliation{$^{1}$Sister Nivedita University, New Town, Kolkata, West Bengal 700156, India\\
$^{2}$School of Basic Sciences, Indian Institute of Technology Bhubaneswar, Jatni, Khurda, Odisha 752050, India}
\vskip-2.8cm
\date{\today}
\vskip-0.9cm

\vspace{5mm}

\begin{abstract}
We briefly review the problem of Brownian motion and describe some intriguing facets. The problem is first treated in its original form as enunciated by Einstein, Langevin, and others. Then, utilizing the problem of Brownian motion as a paradigm and upon using the Langevin equation(s), we present a brief exposition of the modern areas of stochastic thermodynamics and fluctuation theorems in a manner accessible to a non-expert. This is followed by an analysis of non-Markovian Brownian dynamics via generalized Langevin equation(s) in which we particularly shed light onto its derivation, the emergence of the fluctuation-dissipation relation, and the recently-discovered effective-mass framework.
\end{abstract}

\maketitle

\section{Introduction}
Nearly two centuries ago, Robert Brown designed his own
microscope and decided to examine the movement of pollen
particles suspended in a tall jar of water. To his amazement
he saw that the motion of the particles was not regular and occurred in a `zig-zag', random manner even though the jar was
maintained at a constant temperature and pressure \cite{Brown}. The implication was that even in thermal
equilibrium, the pollen particles were apparently in irregular
dynamics which has since been dubbed the Brownian
motion. About eight decades had elapsed before Einstein
gave a theoretical analysis of Brownian motion based on the
existing knowledge of fluid dynamics \cite{Fick,Stokes} and
statistical-thermodynamic concepts due to Maxwell, Boltzmann, and others. The
principal aim of Einstein was to provide a working formula for
the Avogadro number \(N\); in the
process, he succeeded in contextualizing Brown’s observation in terms of Boltzmann’s notion of
thermodynamic fluctuations and diffusion phenomenon \cite{Einstein2}. The paper \cite{Einsteinsumm} presents a summary of Einstein's work on statistical mechanics before 1905 (see also, \cite{SDGResonance2006} for a pedagogic discussion). The papers \cite{history,history1} discuss some historical developments concerning Brownian motion. Shortly after Einstein’s papers on the
subject, Langevin \cite{Langevin} put forward a mathematical framework that
helped elucidate Brownian motion from considerations of
stochastic theory and diffusion processes \cite{zwanbook,SDG_diff,bala}. At a fundamental level, Brownian motion has enlightened us on how
irreversibility occurs in Newtonian and Hamiltonian dynamics which are otherwise invariant under time-reversal \cite{Zwanzig}. Concomitantly, a clear understanding of how a system, when disturbed by external means, eventually arrives at an
equilibrium with the surroundings such as in the linear-response theory of Kubo \cite{Kubo} and Luttinger \cite{Luttinger}, has been
made possible by a proper appreciation of the Brownian
motion.

\vspace{2mm}

What transpired as a curiosity-driven observation by Brown turned out to have a profound
significance for the basic subject of non-equilibrium
thermodynamics and statistical mechanics. So much so that
Kadanoff \cite{Kada}, in his book on statistical mechanics has termed
Einstein’s analysis of Brownian motion as a novel approach to
statistical mechanics, distinct from the Gibbs formalism. The problem of Brownian motion is closely linked to kinetic theory which rests on the Boltzmann (transport) equation. It is now well known that under certain conditions, the Boltzmann equation can directly lead to the Navier-Stokes equations \cite{chapman} which are at the heart of fluid dynamics; this leads to computational techniques involving the so-called lattice Boltzmann method \cite{succi}. Moreover, the Fokker-Planck equation which arises in the Brownian-motion problem can be obtained from the Boltzmann equation \cite{FPB1,FPB2}, tying the problem of Brownian motion closely with the framework of statistical mechanics due to Maxwell, Boltzmann, and Gibbs.

\vspace{2mm}

Besides that, Brownian motion has myriad applications to not
just the physics of fluids but also to: (i) construction industry
and granular matter \cite{Gran}, (ii) dairy industry of casein micelles and
their colloidal-suspension properties \cite{Dairy}, (iii) ecology of aerosol
particles in clouds \cite{aero}, and even to the quantum dynamics of
electrons \cite{bmetal} in metals and semi-metals. In the context of the contemporary focus on nano systems, the adaptation of Brownian motion is a hotly-pursued vocation across disciplines of physics, chemistry,
materials science, and engineering. Quoting Pais \cite{Pais}, ``Einstein
might have enjoyed hearing this, since he was quite fond of
applying physics to practical situations.'' Einstein himself
wrote in 1915, i.e., in the year in which he proposed his general
theory of relativity that ``... the theory of Brownian motion is
of great importance as it permits an exact computation of \(N\).
... The great significance as a matter of principle is, however,
that one sees directly under the microscope a part of the
heat energy in the form of mechanical energy.'' He further
mused in 1917: ``... because of the understanding of the
essence of Brownian motion, suddenly all doubts vanished
about the correctness of Boltzmann’s interpretation of the
thermodynamic laws.''

\vspace{2mm}

Since the detailed theoretical treatment of the problem of Brownian motion \cite{Einstein2,Langevin} (see also, \cite{chandra}), the subject of Brownian motion has experienced considerable growth over the years, including the development of its quantum counterpart \cite{QBM0,QBM01,QBM1,QBM2,QBMfundamental,QBM}. Some recent developments within the domain of classical physics include studies on anomalous diffusion and transport (see for example, \cite{balki,transport1,transport2}), stochastic thermodynamics \cite{sek,sei,res,nano-motor,st10,st11,st12,st13,st14,st15,st16,edgar}, fluctuation theorems \cite{fluc0,fluc1,fluc2,fluc25,fluc3,fluc34,fluc35,fluc4}, active Brownian motion \cite{active1,active2,active3}, non-Markovian dynamics \cite{nonmark0,nonmark,nonmarkovfluid0,nonmarkovfluid1,nonmarkovfluid2,nonmark1,nonmark2,nonmark3,nonmark30}, Brownian functionals \cite{Brownianfunctional}, stochastic resetting \cite{resetting,resetting1}, Brownian motors and engines \cite{engine1,engine2,engine3}, and the list goes on. While it is not practical to review such an enormous volume of developments concerning Brownian motion in a single article, the purpose of this short review is to furnish a quick introduction to the Brownian-motion problem in its original form, together with an emphasis on some recent topics such as stochastic thermodynamics, fluctuation theorems, and non-Markovian Brownian motion. Although the topics such as stochastic thermodynamics and fluctuation theorems which have been developed within the last three decades have undergone considerable development, they can be understood in their simplest forms by analyzing the problem of Brownian motion, the latter being familiar to physicists from different backgrounds. Thus, while the present exposition cannot act as a substitute for a textbook \cite{SDG_diff,bala,zwanbook}, it is hoped that it will serve to familiarize the uninitiated reader to some physical ideas of the field (including some recently-developed ones) while keeping the discussion less technical. 

\vspace{2mm}

Given this background and our motivation behind writing this paper, let us present the organization of the paper. In Sec. (\ref{einsteinsec}), we will provide an overview of Einstein’s
amplification of Brown's observations and the elucidation
of the notion of thermodynamic fluctuations. The
concomitant diffusion processes and their understanding
from a stochastic formalism were carried out by Langevin,
which shall constitute the topic of Sec. (\ref{langevinsec}), followed by a brief discussion on the inclusion of (conservative) potentials in the Langevin equation(s) in Sec. (\ref{potsec}). We will then move on in Sec. (\ref{stochasticsec}) to a new
area in which Brownian motion and the associated Langevin
equation(s) find usefulness in describing fluctuating thermodynamics. This subject, which has been coined the epithet
of stochastic thermodynamics \cite{sek,sei,res} is highly relevant in recent applications \cite{nano-motor,st10,st11,st12,st13,st14,st15,st16,edgar}. After discussing the basic ideas in Sec. (\ref{sec9a}), we shall describe the notion of the Brownian Stirling engine in Sec. (\ref{sec9b}), for which the efficiency in the steady state will be presented both using standard thermodynamics and the Langevin equations. The appearance of irreversibility is discussed specifically in Sec. (\ref{sec9c}), wherein the topic of discussion is fluctuation theorems. This is followed by a brief exposition on a remarkable result called the Jarzynski equality in Sec. (\ref{sec9d}). The Langevin
equation(s) form the basic building block(s) of non-equilibrium
statistical mechanics, not just in clarifying the appearance of
irreversibility, but also in providing practical tools for classical and quantum many-body systems. Moreover, the presence of memory is ubiquitous in various physical problems \cite{nonmark0,nonmark,nonmark1,nonmarkovfluid0,nonmarkovfluid1,nonmarkovfluid2,nonmark2,nonmark3,nonmark30}, a fact that has motivated us to devote the entirety of Sec.  (\ref{nonmarko}) to the treatment of non-Markovian (with memory) Brownian motion. In particular, in Sec. (\ref{nmsec1}), we will introduce the generalized Langevin equation which will be derived from a microscopic system-plus-bath approach in (\ref{nmsec2}). We will then discuss the recently-developed formalism of introducing an effective mass for describing non-Markovian systems with a short memory \cite{nonmark3,nonmark30} in Sec. (\ref{nmsec3}). Finally, in Sec. (\ref{Conc}), we will offer our concluding remarks.

\section{Einstein and Langevin}\label{ELsec}

\subsection{Einstein's treatment of Brownian motion and diffusion}\label{einsteinsec}
In his famous paper \cite{Einstein2} (see also, \cite{Einsteinsumm}) on
Brownian motion, reviewed in \cite{SDGResonance2006},
Einstein makes an ingenious juxtaposition of known results
from gas laws, Stokes' formula for the diffusion constant, and Fick’s law of fluid dynamics \cite{Fick,Fick1} to construct a theory
that allows for a robust estimation of the Avogadro number
\(N\). \textit{Inter alia}, he succeeds to connect Brownian motion with
the diffusion equation, known from Fourier’s work on heat
transport in the early nineteenth century. The salient points are summarized below.
Without loss of generality, we shall confine our treatment to
one dimension \cite{SDGResonance2006}. 

\vspace{2mm}

The experimentally-determined gas laws of Charles and Boyle for an ideal gas can be combined to give the gas equation, which, for one mole of ideal gas reads
\begin{equation}\label{idealgasequation}
PV = RT,
\end{equation}
where \(P\) is the pressure, \(V\) the volume (occupied by one mole of gas), \(T\) the temperature, and \(R\) is the gas constant. When it comes to the
Brownian motion, Einstein invoked the van't Hoff law for the osmotic pressure of a dilute
suspension of solute particles to which equation (\ref{idealgasequation}) applies. Now, considering an
Avogadro number \(N\) of solute particles (\(N\) being a fixed
number), we have \(V = N \nu\), \(\nu\) being
the volume available for each solute particle. We next relate \(\nu\) to \(\rho\), the density of the
solute per unit volume of the suspension \cite{SDGResonance2006} as \(\rho = m/\nu\), where it
is to be noted that \(m\) is the mass of each solute particle, given
by \(M/N\), \(M\) being the molecular weight of the solute. We may also stress that, because Brown's pollen particles, i.e., solutes, float on water, i.e., solvent, their mass densities must be the same (making the `effective' pollen mass vanish; see equation (\ref{13eq}) below) but distinct from `\(\rho\)' defined above\footnote{It may be clarified that because the Brownian solute floats, the
weight of each solute particle is the same as the weight of
displaced solvent, i.e., \(\rho_{\rm solute} \times \nu_{\rm solute} \times g = \rho_{\rm solvent} \times \nu_{\rm (displaced)solvent} \times g\), with \(g\)
being the acceleration due to gravity. But, following Archimedes,
\(\nu_{\rm solute} = \nu_{\rm (displaced)solvent}\), and hence \(\rho_{\rm solute} = \rho_{\rm solvent}\), however, distinct from \(\rho\) introduced above.}. Combining all these, equation (\ref{idealgasequation}) can
be restated as
\begin{equation}\label{Peqnrho}
P = \bigg( \frac{RT}{M}\bigg) \rho.
\end{equation}  However, for Brownian motion to occur there must
be a force \(F\) on the pollen particles due to the cumulative
effect of the collisions with the surrounding fluid molecules. From mechanics, we know that the pressure is the force \(F\) per
unit area. Hence, the elementary pressure across an
elementary length \(dx\) of the solvent is \(dP = F dx/\nu\), where \(\nu\),
from the discussion below equation (\ref{idealgasequation}) can be expressed as \(\nu = M/N\rho\), leading to
\begin{equation}\label{Fmechanics}
F = \bigg(\frac{M}{N \rho}\bigg) \frac{\partial P}{\partial x}.
\end{equation}
 Combining equations (\ref{Peqnrho}) and (\ref{Fmechanics}), we obtain
\begin{equation}\label{Fmechanics1}
F = \bigg(\frac{RT}{N}\bigg) \frac{1}{\rho} \frac{\partial \rho}{\partial x},
\end{equation}
i.e., the force connotes to the density gradient.

\vspace{2mm}

Einstein then resorted to fluid mechanics. The force \(F\) on
the pollen particle is related to its instantaneous velocity \(v\) through
the viscosity \(\eta\) and the radius \(a\) of the particle via Stokes' law \cite{Stokes}
\begin{equation}
F = - 6 \pi \eta a v.
\end{equation} 
On the other hand, \(\rho v\) equals the current
density which is given by Fick's law \cite{Fick1} (analogous to Ohm's law in
electrodynamics in terms of the electrical conductivity), i.e., 
\begin{equation}
\rho v = - D_P   \frac{\partial \rho}{\partial x},
\end{equation}
where, by the suffix `\(P\)' on the diffusion constant \(D\), we
underscore the fact that we are dealing here with ‘physical
diffusion’, arising from the random, zig-zag motion of
Brown’s pollen particles \cite{Narasimhan}. Putting it all together, we arrive at the
desired expression that yields \(N\):
\begin{equation}\label{Davogadroexp}
D_P = \frac{RT}{6 \pi \eta a N} .
\end{equation}
Now comes another innovative step by Einstein. He
implements yet one other result due to Fick on what is called the
diffusion equation. While the physics above (described by equations
(\ref{Peqnrho}) to (\ref{Davogadroexp})) is based on macroscopic concepts, Einstein now
asks: What happens when a single pollen particle is tagged?
Given the observation by Brown that each particle undergoes
a stochastic motion due to the random kicks by the water
molecules, the number of pollen particles per unit volume of
water at a point \(x\) at time \(t\), denoted by \(n(x,t)\), must obey a
diffusion equation
\begin{equation}\label{diffeq}
\frac{\partial n(x,t)}{\partial t} = D_S \frac{\partial^2 n(x,t)}{\partial x^2},
\end{equation}
where the subscript `\(S\)' now stands for `stochastic'. The above-mentioned equation is obtained starting with Fick's first law that the diffusion current of pollen particles is proportional to their concentration gradient, i.e., \(J(x,t) = - D_S \frac{\partial n(x,t)}{\partial x}\), where the proportionality factor \(D_S\) is generically called the transport coefficient. Then, combing this with the local conservation of particle number as dictated by the continuity equation \(\frac{\partial n(x,t)}{\partial t} + \frac{\partial J(x,t)}{\partial x} = 0\) gives us (\ref{diffeq}). The
solution of this equation with ‘free’ boundary conditions is
the well-known Gaussian (see for example, \cite{SDG_diff,bala,zwanbook})
\begin{equation}\label{diffsol}
n(x,t) = \frac{n_0}{\sqrt{4 \pi D_S t}} e^{-\frac{x^2}{4 D_S t}} ,
\end{equation} where \(n_0\) is a normalization constant. Consequently, the mean-squared displacement, given by the second moment of the distribution (\ref{diffsol}) is given by 
\begin{equation}\label{msddiff}
\langle x^2(t) \rangle = 2 D_S t.
\end{equation}
The remarkable insight of Einstein is that as far as Brownian
motion is concerned, the physical diffusion constant \(D_P\) of
the surrounding water medium and the stochastic diffusion
constant \(D_S\) characterizing the stochastic diffusion of the
suspended pollen particles are one and the same, i.e., \(D_P = D_S\). Thus, from equations (\ref{Davogadroexp}) and (\ref{msddiff}), one finds the following fascinating result:
\begin{equation}\label{msdnew}
\langle x^2(t) \rangle = \bigg( \frac{RT}{3 \pi \eta a N} \bigg) t.
\end{equation} 
Equation (\ref{msdnew}) embraces another profound result
contained in the linearity (in time \(t\)) for the mean-squared
displacement which is quite distinct from what is expected
for ballistic motion. Indeed, if one uses Brown’s microscope
to collect several samples of the displacement of a
tagged pollen particle, takes its square and performs an
average over such samples, one arrives at a working formula
for evaluating the value of the Avogadro number \(N\). This was done by Perrin \cite{perrin} whose experiments confirmed the `molecular reality' which earned him the Nobel prize in 1926. 

\vspace{2mm}

The mean-squared displacement is governed by fluctuations
that prevail even in thermal equilibrium (of the surrounding medium, held at a fixed temperature \(T\)). However, the mean-squared displacement also gets related to a dissipative parameter, namely, the viscosity \(\eta\) which impedes the motion
of the pollen particles. Thus, equation (\ref{msdnew}) is a statement of what
we refer to as the first fluctuation-dissipation theorem.
One final remark is in order – while the mean-squared velocity
is time-independent, governed by the equipartition theorem, the
mean-squared displacement is explicitly time-dependent.
Therefore, whereas the velocity is a stationary\footnote{A stochastic process is called stationary if all time-dependent moments are independent of time in the long-time limit. Explicitly, for the present case, one finds for large times, \(\langle v^2(t) \rangle = k_B T/m\), independent of \(t\), while \(\langle x^2(t) \rangle \sim t\), implying that \(x(t)\) is not a stationary process although \(v(t)\) is one.} stochastic
process, the displacement is not! 

\subsection{Langevin equation and recovery of equilibrium results}\label{langevinsec}
As was stressed in Sec. (\ref{einsteinsec}), Einstein’s analysis of Brownian
motion was based on the physical picture of statistical
fluctuations though he alluded to the stochastic nature of
these fluctuations when it comes to diffusion. Just two years
later, Langevin \cite{Langevin} put forward a mathematical framework for the
underlying stochasticity which constitutes the cornerstone of
the modern topic of nonequilibrium statistical mechanics and
open quantum systems. In order to motivate the
Langevin equation(s) we begin with a familiar setup that is
used in elementary physics for the measurement of the
viscosity \(\eta\) of water.

\vspace{2mm}

Imagine that a spherical ball of radius \(a\), mimicking
Brown’s pollen particle, is made to fall freely under gravity
through a tall jar of water, kept in thermal equilibrium at
constant temperature and pressure. The Newtonian equation
of motion for the velocity \(v\), again assumed one-dimensional
for simplicity, reads
\begin{equation}\label{lang1}
m \frac{dv}{dt} = m_{\rm eff} g - 6 \pi \eta a v,
\end{equation}
where \(m\) is the mass of the ball, the buoyancy-corrected
effective mass is
\begin{equation}\label{13eq}
m_{\rm eff} = m \bigg(1 - \frac{\rho_w}{\rho_s}\bigg),
\end{equation} and \(g\) is the acceleration due to gravity. The density \(\rho_w\) is that
of `water' while \(\rho_s\) is that of the `solid' ball. The last
term in equation (\ref{lang1}) is the Stokes force introduced earlier. One important remark is in order here. Because \(v\)
changes sign as \(t \rightarrow - t\), the Stokes term explicitly breaks
the time-reversal invariance of the dynamics, a hallmark of a
dissipative system, outside the realm of usual (conservative) classical and
quantum mechanics. Such a breakdown of time-reversal symmetry is
an inherent feature of Brownian motion, a clear understanding
of which requires the consideration of open systems as will
be discussed in a subsequent section. In the steady-state, when \(v\)
reaches the `terminal velocity' \(v_T\), the left-hand-side of equation
(\ref{lang1}) vanishes, yielding
\begin{equation}
v_T = \frac{m_{\rm eff} g}{6 \pi \eta a}.
\end{equation}
A measurement of \(v_T\) then leads to a scheme for computing \(\eta\). This is a standard high-school experiment albeit the issue
of time-reversal is usually sidetracked. 

\vspace{2mm}

What Langevin did was to interpret equation (\ref{lang1}) as an
`averaged' equation in which the random Brownian motion
of the falling ball is averaged over, denoted by angular brackets below. The underlying framework is – the
random collisions which render the velocity as a stochastic
process, yield a force that has two components: (i) a
systematic one that leads to the Stokes force and (ii) a purely-random one which, upon averaging, must vanish and give rise
to the systematic component. The generalized equations
then read
\begin{equation}\label{1dlangevin}
\frac{dx}{dt} = v, \quad \quad \frac{dv}{dt} + \frac{1}{m} \frac{dV(x)}{dx} = - \gamma v + \Theta(t),
\end{equation} with \(\gamma = 6 \pi \eta a/m\). The second term in the left side of the equation for the velocity is a
generalized version of the Archimedes force (the first term
on the right of equation (12)) in terms of a potential energy \(V(x)\). But a significant additional feature
is the presence of a noise (random force) \(\Theta(t)\), which may only be described by its statistical properties\footnote{The noise assumed here is a special case of a stationary Gaussian process, meaning that its complete statistical description is provided by the first and second moments. All odd moments vanish while the even ones can be expressed as ordered products of second moments. For a one-dimensional stationary Gaussian process, Doob's theorem (see for example, \cite{SDG_diff}) states that the underlying correlation function is an exponential function of the time arguments. The delta function assumed in (\ref{noisestat1}) is a special case.}
\begin{equation}\label{noisestat1}
\langle \Theta(t) \rangle =0, \quad \quad \langle \Theta(t) \Theta(t') \rangle = \frac{2\gamma k_B T}{m} \delta(t-t'). 
\end{equation}
An explanation is in order regarding the interpretation of (\ref{1dlangevin}). The first entry is of course the constitutive definition
of the velocity. The second one is a stochastic differential
equation driven by \(\Theta(t)\), which, upon averaging, yields our
earlier treatment of the viscosity. The second equation in
(\ref{noisestat1}) describes the noise correlation. The
Dirac delta function on the right of (\ref{noisestat1}) suggests that the
correlation vanishes unless \(t=t'\). What is the underlying
meaning of this? Each collision between the ball (\`a la the
pollen particle of Brown) is assumed to be instantaneous, as
in the impact approximation of Maxwellian kinetic theory of
ideal gases. The time of impact -- of electromagnetic origin --
can be estimated to be less than \(10^{-14}\) sec, and hence, in the
context of Brownian motion, can be set to be equal to zero \cite{SDG_diff}. This explains
the presence of the delta function in (\ref{noisestat1}). Next, we note
that the noisy force \(\Theta(t)\) drives the velocity \(v(t)\) to be
stochastic, the correlation of which, discussed below, is
dictated to, by the `Stokes time' \(\gamma^{-1}\) that can be estimated to
be around \(10^{-7}\) sec. This estimate is based on taking the
Brownian particle to be of radius about \(1000\) nm while the
radius of the water molecule is at least three orders of
magnitude smaller \cite{SDG_diff}, making Brownian motion very relevant
for nano, soft-condensed-matter physics, e.g., for colloidal
systems. We will argue below that the Einstein regime which was
covered in the previous section is yet on a longer timescale
(larger than the Stokes time) when the stochasticity of \(x(t)\),
driven by \(v(t)\) (the first entry in (\ref{1dlangevin})) sets in. In summary
then, the validity of the Langevin equation(s), though
restricted to times longer than the impact time, admits of
three distinct stochastic processes, namely, \(\Theta(t)\), \(v(t)\) and finally, \(x(t)\)
on a hierarchy of timescales.

\vspace{2mm}

As in the original problem addressed by Einstein, let us take \(V(x) = 0\). Then the equation satisfied by the velocity reads
\begin{equation}\label{lang2}
\frac{dv}{dt} = -\gamma v + \Theta(t),
\end{equation}
the homogeneous part of which yields an exponentially-decaying solution on the Stokes timescale. Combining this
with the inhomogeneous part driven by \(\Theta(t)\) gives rise to an integral
equation that can be easily manipulated with the aid of the
noise correlation presented in (\ref{noisestat1}) to finally lead to an expression
for the mean-squared velocity as \cite{zwanbook,SDG_diff,bala}
\begin{equation}\label{v2result}
\langle v^2(t) \rangle = v_0^2 e^{-2 \gamma t} + \frac{k_B T}{m} (1 - e^{- 2 \gamma t}),
\end{equation} where \(v_0\) is the velocity at time \(t=0\). Evidently, in the asymptotic region: \(t \gg \gamma^{-1}\), the mean-
squared velocity acquires its equipartition form \(k_B T/m\) known from equilibrium statistical mechanics. This has of course been
built-in, because of our stipulation on the pre-factor in front of the noise correlation function in (\ref{noisestat1}). The fact that the
latter, which determines the stochastic fluctuations of the
random collisions suffered by the Brownian particle, has the
specific form that guarantees attainment of thermal
equilibrium at a rate governed by \(\gamma\), is dubbed the second
fluctuation-dissipation theorem \cite{CW}. 

\vspace{2mm}

We now come to discuss Einstein's diffusive regime. For \(t \gg \gamma^{-1}\), \(dv/dt\) vanishes (we mean the moments of the velocity are independent of time in this regime) and equation (\ref{lang2}) in conjunction with the first of (\ref{1dlangevin}) leads to
\begin{equation}\label{overdampedintro}
\frac{dx}{dt} = \frac{\Theta(t)}{\gamma},
\end{equation}
which can be easily integrated and squared to yield a double
integral of the noise. The use of the noise correlation then
enables us to obtain for large times, the result
\begin{equation}\label{msdlangevin}
\langle x^2(t) \rangle = \bigg(\frac{2 k_B T}{m \gamma}\bigg) t.
\end{equation}
Recalling that \(\gamma = 6 \pi \eta a/m\), the result (\ref{msdlangevin}) is identical to
our earlier-derived result (\ref{msdnew}) for the mean-squared
displacement in the diffusive limit if we note that \(k_B = R/N\). It should be pointed out that the regime for which the above diffusive behavior holds is the so-called `overdamped' regime -- here the moments of the velocity are constant in time, fixed at their equilibrium values (dictated here by the Maxwell distribution). The corresponding equations of motion are therefore termed the overdamped Langevin equations which are obtained by setting \(dv/dt = 0\) in (\ref{1dlangevin}). 

\subsection{Potentials}\label{potsec}
Let us briefly describe the effect of having a nontrivial (but conservative) potential \(V(x)\) on the Brownian dynamics. If a potential is included, a Brownian particle experiences a conservative force equal to \(-\frac{dV(x)}{dx}\), in addition to the environment-induced forces and as a result, the current will have a drift contribution in addition to the diffusion part, i.e., \begin{equation}
J(x,t) = (m \gamma)^{-1} \bigg[-\frac{dV(x)}{dx} n(x,t) - D m \gamma \frac{\partial n(x,t)}{\partial x}\bigg], 
\end{equation} where \(D = k_B T/m\gamma\). Thus, in the equilibrium state, setting \(J(x,t) = 0\) gives us the following time-independent result:
\begin{equation}
n(x) = n_0 e^{- V(x)/k_B T},
\end{equation} independent of \(\gamma\). Therefore, while \(\gamma\) does play a role in determining how fast equilibrium is achieved, it drops out from the equilibrium results; for example, the mean kinetic energy \(\frac{m \langle v^2(t) \rangle}{2}\) depends upon \(\gamma\) (via (\ref{v2result})) but in the long-time limit where equilibrium is reached, the kinetic energy is given by the equipartition theorem. 

\vspace{2mm}

Now, if \(V(x)\) is a confining potential such as a harmonic trap, a Brownian particle moving through it is `trapped' and cannot diffuse in the way a `free' Brownian particle can. For example, in the harmonic-oscillator case, i.e., with \(V(x) = \frac{m \omega_0^2 x^2}{2}\), the position-correlation function for equal times takes the following form in the long-time limit \cite{chandra}: 
\begin{equation}\label{oscmsp}
\langle x^2(t) \rangle = \frac{k_B T}{m \omega_0^2},
\end{equation} which is a constant and does not increase with \(t\). On the other hand, if \(V(x)\) has multiple minima with barriers (local maxima) between them, the environment-induced fluctuations make it possible for Brownian particles to cross the barrier to move from one minimum to another. This feature is quantified by Kramer's escape rate \cite{chandra} which has also been formulated in quantum-mechanical settings (see \cite{eli} and references therein). However, it must be mentioned that for periodic potentials, one can still obtain a diffusive behavior of the Brownian particle \cite{history,period}. 

\vspace{2mm}

One ought to notice that the occurrence of a potential introduces at least one new timescale into the problem. For simplicity, let us consider the harmonic potential \(V(x) = \frac{m \omega_0^2 x^2}{2}\) for which the potential introduces a new timescale \(\omega_0^{-1}\) which may be completely different from the timescale \(\gamma^{-1}\), associated with the Stokes force exerted on a Brownian particle due to the surrounding fluid. In this case, the dynamics \(x = x(t)\) involves two modes with frequencies (see \cite{chandra} for example)
\begin{equation}
\omega_\pm = - \frac{\gamma}{2} \pm \frac{1}{2} \sqrt{\gamma^2 - 4 \omega_0^2},
\end{equation} i.e., 
the solution of the corresponding Langevin equation turns out to be
\begin{equation}\label{xtoscillator}
x(t) = \frac{1}{\omega_+ - \omega_-} \int_0^t dt' \left[ e^{\omega_+(t-t')} - e^{\omega_-(t-t')} \right] \Theta(t'),
\end{equation}
where we have discarded the initial condition as we are interested in the long-time behavior. From (\ref{xtoscillator}) follows (\ref{oscmsp}) in the long-time limit if one makes use of the noise correlation given in (\ref{noisestat1}). 

\section{Stochastic thermodynamics}\label{stochasticsec}
Standard thermodynamics, considered as the mother of all sciences – physics, chemistry, and biology -- pertains to a system in equilibrium. The underlying import of the clich\'e `equilibrium' is that all fields such as the temperature, the pressure, the magnetic field, the electric field, etc., are held fixed and the system is allowed sufficient time so that all dynamical variables, e.g., position, momentum, and their functions attain values that are on the average constant in time. If any of the above-mentioned fields is changed to another value, the system in general is expected to come to a new equilibrium after a time longer than what is known as the `relaxation time', akin to the inverse of the friction coefficient $\gamma$ appearing in a Langevin equation for Brownian motion as discussed earlier. Standard thermodynamics, however, does not touch upon the origin of the relaxation time or the time evolution of the system. On the other hand, when we come to systems that are so tiny such as nano systems that they are hardly ever in equilibrium, there is a need to go beyond standard thermodynamics and treat fluctuating, time-dependent effects.

\vspace{2mm}

Given this background, and also the axiom that (microscopic) statistical mechanics provides the base for (macroscopic) thermodynamics, a natural question may arise: Can the time-dependent Langevin equations be effectively utilized to formulate a theory for a `time-dependent thermodynamics'? Such time-dependencies would evidently encompass fluctuations in thermodynamic quantities when the system is far from equilibrium. Thus, these fluctuations are very distinct from fluctuations around an equilibrium, which can be probed through `linear response'. Usually, fluctuations in thermodynamic variables are considered `small' for macroscopic objects. However, recently, there has been an upsurge of interest in observing the behavior of nanoscale systems, especially in biology, comprising cell motion, biopolymers, colloidal suspensions, molecular motors, `active' matter, nano engines, and so on. While the aforesaid systems belong to the realm of classical thermodynamics, there are issues in solid state systems of nano junctions, q-bits etc., classified as mesoscopics, that require a generalization of equilibrium thermodynamics to the non-equilibrium and quantum regimes (see for example, \cite{st10,st11,st12,st13,st14,st15,st16}). The above considerations then motivated, first Sekimoto~\cite{sek} and subsequently, Seifert~\cite{sei} to put forward what they called `stochastic thermodynamics' (see \cite{res} for a pedagogic discussion). Because the Langevin equations subsume the stochastic fluctuations induced by the environment, they are omnipresent in the basic building blocks of stochastic thermodynamics. Below, let us describe some basic ideas associated with this formalism. 

\subsection{Basic ideas}\label{sec9a}
Recall that the first two laws of thermodynamics can be combined to write the incremental work, specified by the symbol \(\Delta\) prefixing the work \(W\) as
\begin{equation}\label{fl9}
\Delta W = TdS - dE = - dF - S dT,
\end{equation} where \(T\) is the temperature, \(S\) is the entropy, \(E\) is the mean energy, and \(F\) is the Helmholtz free energy. If we examine equation~(\ref{fl9}), it is evident that the work done by, say an engine, can be computed by simply integrating \(dE\) over an isentropic process or by integrating \(dF\) over an isothermal process. It should be added that \(\Delta W\) itself is determined by \(P dV\) (for mechanical work), \(-B dM\) (for magnetic work), \(\mathcal{E} d\Pi\) (for electrical work), and so on. Here \(P\) is the pressure, \(V\) is the volume, \(B\) is the magnetic field, \(M\) is the magnetization, \(\mathcal{E}\) is the electrical field, and \(\Pi\) is the electric polarization.

\vspace{2mm}

With this brief summary of standard thermodynamics, let us revisit the classical Langevin equation with constant friction \(\gamma\), equation~ (\ref{lang2}), with an additional potential term, i.e.,
\begin{equation}\label{classlangstoc}
\frac{dv}{dt} = - \frac{1}{m} \frac{dV(x)}{dx} - \gamma v + \Theta(t).
\end{equation}  The Sekimoto stratagem is to assume that the state of the system is altered by a small amount \(dx\). Multiplying the forces in equation~(\ref{classlangstoc}) by (\(-dx\)) would then represent the energy balance:
\begin{equation}\label{EBstoc1}
-\bigg(\frac{dv}{dt}\bigg) dx = \frac{1}{m} \bigg(\frac{dV(x)}{dx}\bigg) dx - \big[-\gamma v + \Theta(t)\big] dx.
\end{equation}

\vspace{2mm}

We may now manipulate the left-hand side of equation~(\ref{EBstoc1}) and rewrite it for incremental changes as
\begin{equation}\label{deltap1}
- m \Delta v \bigg(\frac{\Delta x}{\Delta t}\bigg) = - d (mv^2/2).
\end{equation}
Imagine at this stage that the potential energy \(V(x)\) depends on not just the coordinate of the system but also on another parameter `\(\epsilon\)' that can be independently varied. Then,
\begin{equation}
dV(x,\epsilon) = \bigg(\frac{\partial V(x,\epsilon)}{\partial x}\bigg)_\epsilon dx +  \bigg(\frac{\partial V(x,\epsilon)}{\partial \epsilon}\bigg)_x d\epsilon,
\end{equation}
which implies that the first term on the right-hand side of equation~(\ref{EBstoc1}) will have to be modified to \(dV(x,\epsilon) - (\partial V(x,\epsilon)/\partial \epsilon)_x d\epsilon\), thus yielding from equations~(\ref{EBstoc1}) and~ (\ref{deltap1}) that
\begin{equation}\label{dHstoc1}
0 = d\mathcal{E} - \bigg(\frac{\partial V(x,\epsilon)}{\partial \epsilon}\bigg)_x d\epsilon - m \big[-\gamma v + \Theta(t)\big] dx,
\end{equation} where \(\mathcal{E}\) is the (fluctuating) mechanical energy 
\begin{equation}
\mathcal{E} = \frac{mv^2}{2} + V(x,\epsilon) .
\end{equation}
Given that the energy function \(E\) connotes to the `statistical' average of \(\mathcal{E}\), i.e., \(E = \langle \mathcal{E} \rangle\), equation~(\ref{dHstoc1}) can be re-interpreted as the law of thermodynamics vis \`{a} vis equation~(\ref{fl9}) if we recognize that \(- (\partial V(x,\epsilon)/\partial \epsilon)_x d\epsilon\) is the incremental work done by the system. Thus, we reach the following identification as far as stochastic thermodynamics is concerned:
\begin{equation}\label{fl10}
\Delta W = \Delta Q - d\mathcal{E},
\end{equation} where we have
\begin{eqnarray}
d\mathcal{E} &=& d(mv^2/2 + V(x,\epsilon)), \label{dEstoc} \\
\Delta W &=& - \bigg(\frac{\partial V(x,\epsilon)}{\partial \epsilon}\bigg)_x d\epsilon, \label{dWstoc}\\
\Delta Q &=& m \big[-\gamma v + \Theta(t)\big] dx . \label{dQstoc}
\end{eqnarray}
The basic thermodynamic principle is thus nicely encapsulated under equation~(\ref{fl10}); the useful work done by the system is the heat dumped into it by the external bath minus the increase in its mean energy. It must be underscored, however, that unlike standard thermodynamics, all the quantities appearing on the right-hand sides of equations~(\ref{dEstoc}) -- (\ref{dQstoc}) are fluctuating, stochastic processes, which would have to be treated with the aid of the underlying probability distributions. 

\subsection{Harmonic oscillator as a classical nano engine}\label{sec9b}
We envisage our system to be a nano harmonic oscillator of mass \(m\) and a frequency \(\omega_0\) that can be varied with time. The potential energy is given by
\begin{equation}
V(x,\epsilon) = \frac{\epsilon x^2}{2} = \frac{m \omega_0^2 x^2}{2},
\end{equation} where we have \(\epsilon = m\omega_0^2\). The standard thermodynamic scenario is that the particle endowed with a constant frequency \(\omega_0\) is kept in contact with a heat bath comprising a large system of many degrees of freedom and in equilibrium at a constant temperature \(T\). In stochastic thermodynamics, however, we will take \(T\), just as \(\omega_0\), to be also time-dependent. The scales on which these variations take place would yield different results in different regimes of experimental timescales at hand. Before proceeding further, we summarize the well-known results for the thermodynamic potentials, which can be derived from the canonical ensemble:
\begin{eqnarray}
E = k_B T, \quad \quad F = k_B T \ln \bigg(\frac{\hbar \omega_0}{k_B T}\bigg), \\
S = k_B \bigg[1 -  \ln \bigg(\frac{\hbar \omega_0}{k_B T}\bigg) \bigg] . \nonumber
\end{eqnarray}
Evidently, \(S\) depends on the temperature only through the dimensionless ratio \(\alpha = \hbar \omega_0/k_BT\), which must be kept fixed for an isoentropic process; hence, if \(\omega_0\) has a time variation, so must \(T\) have. Further, in standard thermodynamics, we have
\begin{equation}
dF = - SdT - PdV,
\end{equation} which signifies that \(F\) should be regarded as a function of two variables \(T\) and \(V\). It is then appropriate to say that a variation in \(\omega_0\) is tantamount to a change in \(V\); conversely, an isochoric process is the one in which \(\omega_0\) is kept fixed. In what follows, we will apply the formalism developed here to calculate the efficiency of a Stirling engine, first from standard thermodynamics and then from stochastic thermodynamics in the stationary state (that is, using equilibrium results), in order to have a consistency check. A schematic of the Stirling cycle is shown below.\\

\noindent
\begin{align} \nonumber
&\begin{array}{cccccc}
\omega_2,T_h~&  &\text{Isothermal}& &&\omega_1,T_h\\
& 1 &\longrightarrow& 2&&\\
&   &&\\
 \text{Isochoric}&\uparrow&&\downarrow&&\text{Isochoric}\\
&   &&\\
& 4&\longleftarrow&3 &&\\
\omega_2,T_c~& &\text{Isothermal}& &&\omega_1,T_c\\
&&&&&\\ \end{array} \\
&~~~~~~~~~~~~~~~\omega_2>\omega_1,~T_h>T_c~~~~~~~~~~
\nonumber
\end{align}

In the calculations involving the Stirling cycle, while referring to specific values of the trap frequency \(\omega_0\) at points `1' and `2', we will drop the subscript `0', and use the notation \(\omega_1\) and \(\omega_2\), respectively. Now, the net work done is given by
\begin{equation}\label{W1234}
W = W_{1 \rightarrow 2} + W_{3 \rightarrow 4}.
\end{equation}
Referring to the discussion below equation~(\ref{fl9}), it is clear that
\begin{eqnarray}
W &=& -[F(\omega_1,T_h)-F(\omega_2,T_h)] -[F(\omega_2,T_c) -F(\omega_1,T_c )] \nonumber \\
&=& k_B (T_h – T_c) \ln \bigg(\frac{\omega_2 }{ \omega_1}\bigg). \label{Wstoc11}
\end{eqnarray}
On the other hand, the heat transferred from the hot source is given by
\begin{eqnarray}
Q &=& [E_{1 \rightarrow 2} + E_{4 \rightarrow 1}] + W_{1 \rightarrow 2} \nonumber \\
&=& 0 + k_B (T_h -T_c) + k_B T_h \ln \bigg(\frac{\omega_2 }{ \omega_1}\bigg). \label{Qstoc11}
\end{eqnarray}
Combining equations (\ref{Wstoc11}) and (\ref{Qstoc11}), we obtain for the efficiency of the Stirling cycle, the following result:
\begin{equation}\label{stirlingeff1}
\eta^{(\mathrm{Stirling})} = \frac{W}{Q} = \eta^{(\mathrm{Carnot})} \bigg[1 + \frac{\eta^{(\mathrm{Carnot})}}{\ln (\omega_2/\omega_1)} \bigg]^{-1},
\end{equation} where \(\eta^{(\mathrm{Carnot})} = 1 - T_c/T_h\) is the Carnot efficiency.

\vspace{2mm}

We now turn to the computation of efficiency from stochastic thermodynamics and demonstrate that in the stationary state, the answer matches with equation~(\ref{stirlingeff1}). The incremental work operator is now given by
\begin{equation}
\Delta W = -m \omega_0 x^2 d\omega_0.
\end{equation} From equation~(\ref{W1234}), we have
\begin{eqnarray}
W &=& - \int_{\omega_1}^{\omega_2} d\omega \frac{\langle m \omega^2 x^2 \rangle_{T_c}}{\omega}  + \int_{\omega_1}^{\omega_2} d\omega \frac{\langle m \omega^2 x^2 \rangle_{T_h}}{\omega}  \nonumber \\
&=& k_B (T_h - T_c) \ln \bigg( \frac{\omega_2}{\omega_1}\bigg), \label{Wstoc22}
\end{eqnarray} where we have employed the classical equipartition theorem in extracting the angular-bracketed quantities in the integrands. As expected, this result agrees with equation (\ref{Wstoc11}). Next, in computing the heat gained from the hot source in the stationary state, we need \(E_{1 \rightarrow 2} + E_{4 \rightarrow 1}\) (see equation (\ref{Qstoc11})). We have
\begin{equation}
E_{1 \rightarrow 2} = \int_1^2 d\bigg(\frac{m \langle v^2\rangle}{2} + \frac{m \omega_0^2 \langle x^2 \rangle}{2}\bigg) = 0,
\end{equation} having used the equipartition theorem again and the fact that the temperature is the same at points `1' and `2'. Similarly, we find
\begin{equation}
E_{4 \rightarrow 1} = \int_4^1 d\bigg(\frac{m \langle v^2 \rangle}{2} + \frac{m \omega_0^2 \langle x^2 \rangle}{2}\bigg) = k_B (T_h - T_c).
\end{equation}
Thus, we have
\begin{eqnarray}
Q &=& [E_{1 \rightarrow 2} + E_{4 \rightarrow 1}] + W_{1 \rightarrow 2} \nonumber \\
&=& k_B (T_h - T_c)  + k_B T_h \ln \bigg( \frac{\omega_2}{\omega_1}\bigg),
\end{eqnarray} where \(W_{1 \rightarrow 2}\) has already been calculated in equation~(\ref{Wstoc22}). Once again, this agrees with the expression in equation (\ref{Qstoc11}), confirming that the stationary-state result for the efficiency in stochastic thermodynamics must conform to that found from standard thermodynamics. Summing up, the subject of stochastic thermodynamics takes us beyond standard thermodynamics and hence incorporates in a natural manner the dissipative parameters. \textit{Inter alia}, it provides a framework in which thermodynamic potentials like the mean energy and the free energy are given a new meaning in terms of dynamics. The appropriate dynamical equation is the Langevin equation, which is at the heart of the Brownian motion. 

\subsection{Fluctuation theorems}\label{sec9c}
In the preceding discussion, we considered working with an `energy balance' condition, as in (\ref{fl10}). Notice that in (\ref{fl10}), the quantities that appear are fluctuating and only when averaged over the noisy effects inherent in the Langevin equation(s), can they be regarded as constituting a first law of equilibrium thermodynamics. In a sense, therefore, (\ref{fl10}) is more general than equilibrium thermodynamics as it includes the fluctuating effects. This brings us to the closely-related area of fluctuation theorems \cite{fluc0,fluc1,fluc2,fluc25,fluc3,fluc34,fluc35,fluc4}. 

\vspace{2mm}

Consider a generic (say, isolated) thermodynamic system which is not in equilibrium. Let us ask: in a given amount of time \(\tau\), should the entropy increase or decrease? Obviously, our common sense stemming from years of learning standard thermodynamics tells us that the entropy should increase. It could, however, decrease without violating the first law of thermodynamics but the second law does not allow this. Although thermodynamics, which is based on empirical evidence, does advocate for the non-decreasing nature of entropy for an isolated system, it does not tell us why that should happen given that the motion of the constituents of a gas (say) is governed by Newton's equations which are reversible (consider elastic collisions between gas molecules). With statistical mechanics as the foundation for thermodynamics, it seems only natural to ask for a microscopic meaning of the second law of thermodynamics based on statistical concepts. In this regard one encounters the notion of a fluctuation theorem: while entropy production could be both positive and negative, the probability of negative entropy production (i.e., decrease in entropy) is `suppressed' over positive entropy production. In the limit of equilibrium for macroscopically-large systems, the probability of negative entropy changes becomes practically zero and this leads to the second law of thermodynamics. Thus, the fluctuation theorem implies that although entropy could `in-principle' decrease, it is less likely to happen, therefore, lending a statistical interpretation to the second law of thermodynamics. We must reiterate that this does not say that the second law of thermodynamics is wrong, for the second law applies for macroscopic systems in which case the fluctuation theorem gives identical results. However, when one is considering nano-sized systems, such fluctuations cannot be ignored and can have strong implications. 

\vspace{2mm}

Before taking up any calculations, let us make the preceding discussion more precise. It should be emphasized upon that if the equations of motion or their solutions (the phase-space trajectories) of a system are known, then that determines whatever can be known about the system. However, the equations of motion are time-reversible, meaning that for every trajectory, there exists a time-reversed trajectory or ``anti-trajectory'' which also solves the same equations of motion. Then, the relative probabilities of observing `bundles' of conjugate trajectories can be used to quantify the ``macroscopic reversibility'' of the system, i.e., if the probabilities of observing all the allowed trajectories and their time-reversed counterparts are equal, then one can conclude in favor of reversibility. However, if the relative probability of observing the anti-trajectories is small (vanishingly, for macroscopic systems), then one can conclude that the system admits irreversibility. The fluctuation theorem in the form presented by Evans and Searles \cite{fluc0} describes such relative probabilities of observing trajectories over a certain duration \(\tau\) as described by a so-called `dissipation function' \(\Omega_\tau\), taking arbitrary values \(\mathcal{A}\) and \(-\mathcal{A}\), as
\begin{equation}\label{EvansSearlesFluc}
\frac{p(\Omega_\tau = -\mathcal{A})}{p(\Omega_\tau = \mathcal{A})} = e^{-\mathcal{A}},
\end{equation} where \(p(\Omega_\tau = \mathcal{A})\) is the probability of forward evolution while \(p(\Omega_\tau = -\mathcal{A})\) denotes the same for time-reversed evolution. The result (\ref{EvansSearlesFluc}) demonstrates the exponential suppression of trajectories corresponding to \(\Omega_\tau=-\mathcal{A}\), demonstrating the `statistical' emergence of irreversibility. In what follows, we will consider the Brownian-motion problem and will formulate a version of fluctuation theorem in the steady state, attributed to Gallavotti and Cohen \cite{fluc1}. This will present us with a flavor of the fluctuation theorems using the simple framework used in Sec. (\ref{ELsec}) in which the Langevin equation shall play an important role. 

\vspace{2mm}

Let a Brownian particle be acted upon by an external force \(f(t)\) -- say, the particle is electrically charged and one has switched on a time-dependent electric field. Then, the work done on the particle over a period of time \(t\) turns out to be 
\begin{equation}\label{workdeforiginal}
W(t) = \int_0^t v(t') f(t') dt',
\end{equation}
where \(v(t)\) is dictated by Langevin dynamics. 
\begin{widetext}
Since \(v(t)\) for Brownian motion is a Gaussian process (because \(\Theta(t)\) is Gaussian), so is \(W(t)\). Thus, the characteristic function \(C(h) = \langle e^{\mathrm{i}h W} \rangle\) can be decomposed as
\begin{eqnarray}
C(h) &=& e^{\mathrm{i}h \langle W \rangle} \langle e^{\mathrm{i} h \int_0^t dt'~f(t') [v(t') - \langle v(t') \rangle]} \rangle \nonumber \\
&=& e^{\mathrm{i}h \langle W \rangle}  e^{ - \frac{h^2}{2} \int_0^t dt'~\int_0^t dt''~f(t') f(t'') [\langle v(t') v(t'') \rangle - \langle v(t') \rangle \langle v(t'') \rangle]} , \label{x8}
\end{eqnarray}
\end{widetext}
where, in the last step, we have used the property of Gaussian processes as far as their cumulants are concerned \cite{SDG_diff} and have written
\begin{equation}
\langle W \rangle = \int_0^t f(t') \langle v(t') \rangle dt',
\end{equation} where we have also suppressed the time argument from the left-hand side with the understanding that we are considering the long-time limit. Next, defining the variance \(\sigma_W^2\) by the exponent in the second term in~(\ref{x8}), the characteristic function can be written as
\begin{equation}
C(h) = e^{\mathrm{i}h \langle W \rangle} e^{\frac{-h^2 \sigma_W^2}{2}}.
\end{equation}
Thus, performing a Fourier transform of \(C(h)\), one finds that \(p(W)\) is also a Gaussian, i.e., 
\begin{equation}
p(W) = \frac{1}{\sqrt{2 \pi \sigma_W^2}} e^{-\frac{(W - \langle W \rangle)^2}{2 \sigma_W^2}},
\end{equation} from which follows that
\begin{equation}\label{GCFT}
\frac{p(-W)}{p(W)} = e^{-\alpha W }, \quad \quad \alpha = \frac{ 2 \langle W \rangle}{\sigma_W^2}.
\end{equation}
The statement (\ref{GCFT}) is the celebrated Gallavotti-Cohen statement of the fluctuation theorem \cite{fluc1,fluc25,fluc35}. A direct calculation from the Langevin equation reveals that the variance of work in the long-time limit turns out to be \(\sigma_W^2 = 2 k_B T \langle W \rangle\), revealing that \(\frac{p(-W)}{p(W)} = e^{- W /k_B T }\). Physically, the probability of `negative work' is exponentially suppressed! This justifies the microscopic irreversibility of the Brownian-motion problem. 

\subsection{Jarzynski equality}\label{sec9d}
Let us now come to the remarkable result due to Jarzynski \cite{fluc2} in the context of the stochastic approach to thermodynamics which has now come to be called the Jarzynski equality. Let us take a system which is initially at equilibrium with temperature \(T\). It is then `transiently' disturbed over some limited interval of time. This disturbance will be associated with some external work, let's call it \(W\); it may leave the system in some new equilibrium state. As discussed earlier, since the microscopic state of the system is probabilistic, the work \(W\) will have an ensemble of realizations although it also depends on the nature of the disturbance imposed upon the system and part of it (or perhaps all of it) could be dissipated as heat. One could now ask: is \(W\) related to the equilibrium state that the system eventually reaches? The answer is provided by the Jarzynski equality which states that
\begin{equation}\label{JE}
\langle e^{-W/k_B T} \rangle = e^{-\Delta F/k_B T},
\end{equation} where \(\Delta F\) is the change in the free energy between the initial and final equilibrium states (the latter may occur at infinite time). Thus, (\ref{JE}) is a remarkable statement relating transient irreversible processes to infinite-time differences between free energies. The averaging on the left-hand side is performed over the ensemble of \(W\) realizations. The fluctuation theorems may therefore be viewed as appropriate generalizations of the Jarzynski equality. 

\vspace{2mm}

As a simple illustration of how the equality (\ref{JE}) may be applied, let us consider the problem of Brownian motion of a suspended particle in a fluid medium as discussed in some detail in the preceding sections (see \cite{peda} for a pedagogic approach). Consider the scenario taken in Sec. (\ref{sec9c}) for the derivation of the Gallavotti-Cohen fluctuation theorem, with a perturbing force \(f\) whose details we will specify later. Thus, over a displacement \(\Delta x\), the work done is
\begin{equation}\label{Wdefinitionfx}
 W = f \Delta x.
\end{equation}
Moreover, in this situation the entire work done is dissipated leading to \(\Delta F = 0\), i.e., the initial and final (equilibrium) states are identical. Thus, the equality (\ref{JE}) implies that 
\begin{equation}\label{oneWork}
\langle e^{- W/k_B T} \rangle = 1.
\end{equation} One can evaluate (\ref{oneWork}) by applying the cumulant expansion as done in (\ref{x8}) earlier. Because the process is Gaussian, this gives
\begin{equation}
e^{- (1/k_B T) \langle W \rangle + (1/2k_B^2 T^2) \sigma_W^2} \approx 1,
\end{equation} implying that the variance is \(\sigma_W^2 = 2 k_B T \langle W \rangle\), a result that can also be obtained using Langevin equations in the long-time limit as remarked below (\ref{GCFT}). We will now assume up to a first approximation that the inclusion of \(f\) does not affect the diffusive behavior of the Brownian particle, i.e., we can still write \(\sigma_x^2 \approx 2 D \Delta t \) for a time increment \(\Delta t\), where \(D\) is the diffusion constant introduced earlier. Thus, \(\sigma_W^2 = f^2 \sigma_x^2 = f^2 \times (2 D \Delta t)\). Using \(\sigma_W^2 = 2 k_B T \langle W \rangle\) as justified earlier, one finds by equating the two expressions for \(\sigma_W^2\) that
\begin{equation}
 k_B T \frac{\langle W \rangle}{\Delta t} = f^2 D. 
\end{equation} Upon using (\ref{Wdefinitionfx}), one gets 
\begin{equation}
k_B T \frac{\langle \Delta x \rangle}{\Delta t} = f D. 
\end{equation}Finally, defining the force \(f\) to be equal and opposite to the drag force, i.e., \(f = m \gamma v = m \gamma (\langle \Delta x \rangle/\Delta t)\), one immediately obtains
\begin{equation}
D = \frac{k_B T}{m \gamma}, 
\end{equation} a result which was obtained earlier in (\ref{msdnew}) and (\ref{msdlangevin}). Thus, the Jarzynski equality (\ref{JE}) naturally leads to the fluctuation-dissipation relation. It may be remarked that when one enters the realm of fluctuation theorems and the Jarzynski equality, there is a subtle crossover from nonequilibrium (stochastic) stationary dynamics to generalized processes that are not-necessarily stationary.

\section{Beyond Markovian processes}\label{nonmarko}
So far we have discussed Brownian motion with two special properties, namely, (i) that the Stokes force experienced by a Brownian particle depends only upon the instantaneous velocity and not upon past velocities, (ii) that the collision-impact time between the Brownian particle and a fluid particle is zero (or infinitesimally small) leading to a noise that is delta-correlated as in (\ref{noisestat1}). While these two properties which have to do with lack of memory (called `Markovian-ness') hold reasonably well in the original version of the Brownian-motion problem, there are instances in fluid mechanics \cite{nonmarkovfluid0,nonmarkovfluid1,nonmarkovfluid2} (often referred to as the Boussinesq–Basset memory force), active Brownian motion (see for example, the paper \cite{active3}), delay-differential equations \cite{delay} where memory effects are important. For example, consider the setup of \cite{active3} in which one considers an active Brownian particle on a plane with a fluctuating orientational angle. The corresponding Langevin equations take the following overdamped form: 
\begin{equation}
\frac{dx}{dt} = v_0 \cos \phi(t), \quad \frac{dy}{dt} = v_0 \sin \phi(t), \quad \dot{\phi} = \sqrt{2D_R} \eta(t),
\end{equation} where \(\langle \eta(t) \eta(t') \rangle = \delta(t-t')\) and \(D_R\) is the rotational diffusion constant. In this case, one can define two `effective' noises \(\Theta_x(t) = \cos \phi(t)\) and \(\Theta_y(t) = \sin \phi(t)\), which are not only bounded as \(|\Theta_{x,y}(t)| \leq 1\) \(\forall\) \(t\) (unlike the noise presented in (\ref{noisestat1})) and are mutually-correlated, but are also non-Markovian as for large times (with \(t-t'\) finite), one has \cite{active1}
\begin{equation}
\langle \Theta_x(t) \Theta_x(t') \rangle \sim e^{-D_R |t-t'|},
\end{equation} and similarly for \(\Theta_y(t)\). Note that \(D_R\) introduces a new timescale such that for \(|t-t'| \gg D_R^{-1}\), one obtains delta-correlated noises in the \(x\) and \(y\) directions, recovering the diffusive behavior of standard Brownian motion as discussed earlier. However, generally the behavior of the mean-squared displacement is significantly different in this scenario, with the emergence of anisotropic behavior; the mean-squared displacement in the \(x\)-direction is much smaller than that in the \(y\)-direction \cite{active1}. 

\subsection{Generalized Langevin equation}\label{nmsec1}
Preserving the linearity of the equations of motion, when non-Markovian Brownian dynamics is involved, it can be described (at least formally) by `generalized' Langevin equations (once again, restricting ourselves to one dimension) \cite{Zwanzig}
\begin{equation}\label{genlang}
\frac{dx}{dt} = v, \quad \frac{dv}{dt} = - \int_0^t \gamma(t-t') v(t') dt' + \Theta(t),
\end{equation} where \(\gamma(t)\) is called the dissipation/damping/friction kernel which is defined to vanish for negative arguments so as to respect the principle of causality and \(\Theta(t)\) is not-necessarily delta-correlated. 

\vspace{2mm}

Recall that the notion of delta-correlated noise arises from instantaneous collisions between the Brownian particle and the fluid molecules. While the instantaneity of a collision is a standard approximation in kinetic theory, in several realistic situations, there is a typical `collision timescale' which can be attributed to the fact that molecules are not exactly `hard spheres' but rather exert a short-ranged repulsive force when two particles collide. Thus, the random force \(\Theta(t)\) experienced by the Brownian particle is only approximately delta-correlated; however, this is often an extremely-good approximation in typical situations with colloidal (Brownian) particles suspended in liquids. Let us assume that there is a timescale (however small) associated with the random noise \(\Theta(t)\), i.e., we can write the correlation function \(\langle \Theta(t) \Theta(t')\rangle = K(t-t')\), where we have assumed that the correlations are time-homogenous and \(\tau_c\) is a newly-introduced timescale such that \(\lim_{\tau_c \rightarrow 0}K(t-t') \propto \delta(t-t')\). Going by our previous understanding that the damping force also originates from the above-mentioned collisions, one is immediately led to believe that \(\gamma(t-t')\) is generally not a delta function of its argument, giving us a generalized Stokes force featuring in (\ref{genlang}). Since the Brownian particle reaches a state of thermal equilibrium (in the sense that its kinetic energy is given by equipartition theorem, as consistent with a Boltzmann-Gibbs distribution), it is not hard to convince oneself that \(\gamma(t-t')\) must be related to \(K(t-t')\); this is just the fluctuation-dissipation relationship. The question that we now ask is: what are the specifics of such a relationship given that we have not specified what \(K(t-t')\) and \(\gamma(t-t')\) are. In the following, we will provide an answer to this question by analyzing a microscopic approach (see \cite{nonmark0} for a derivation of the fluctuation-dissipation relation in a generic setting).

\subsection{System-plus-bath approach}\label{nmsec2}
The equation (\ref{genlang}) may be derived starting with a system+bath approach, which will also allow us to examine whether the fluctuation-dissipation theorem gets modified due to the non-Markovian nature of the Brownian dynamics. Typically, a good (and analytically tractable) description of a heat bath is to consider it to be a large bunch of harmonic oscillators. Of course, the number of degrees of freedom would have to be very large or otherwise one would hit Poincar\'e recurrences\footnote{For the benefit of the non-expert reader, it may be pointed out that in Hamiltonian dynamics, the Poincar\'e recurrence theorem states that a system after a sufficiently long but finite time (called the recurrence time), returns to a state that is arbitrarily close to the initial state. However, if the number of degrees of freedom is large enough, the recurrence time is long enough for the emergence of `irreversibility' for practical purposes.}. Thus, keeping the above in mind, we would take the Hamiltonians of the system and the bath to be the following \cite{Zwanzig,FKM,Ull,mag}:
\begin{equation}
H_{\rm System} = \frac{p^2}{2m}, \quad \quad H_{\rm Bath} = \sum_{j=1}^N \bigg[ \frac{p_j^2}{2m_j} + \frac{ m_j \omega_j^2 q_j^2}{2} \bigg],
\end{equation} where the heat bath is taken to be composed of \(N\) independent harmonic oscillators and we will take \(N\) to be large (theoretically, \(N \rightarrow \infty\)). Here, \((x,p)\) and \((q_j,p_j)\) are the position and momentum variables of the system and the \(j\)-th oscillator of the heat bath, respectively. Now comes the issue of determining how the system and the bath interact; the simplest scheme would be via bilinear coupling, i.e., say, the system's coordinate is coupled with that of the bath oscillators. This suggests an interaction Hamiltonian of the form \(H_{\rm Int} = - x \sum_{j=1}^N c_j q_j\), where the real constants \(\{c_j\}\) are the system-bath microscopic-coupling constants and we have included the negative sign for later convenience. 

\vspace{2mm}

\begin{figure}
\begin{center}
\includegraphics[scale=0.47]{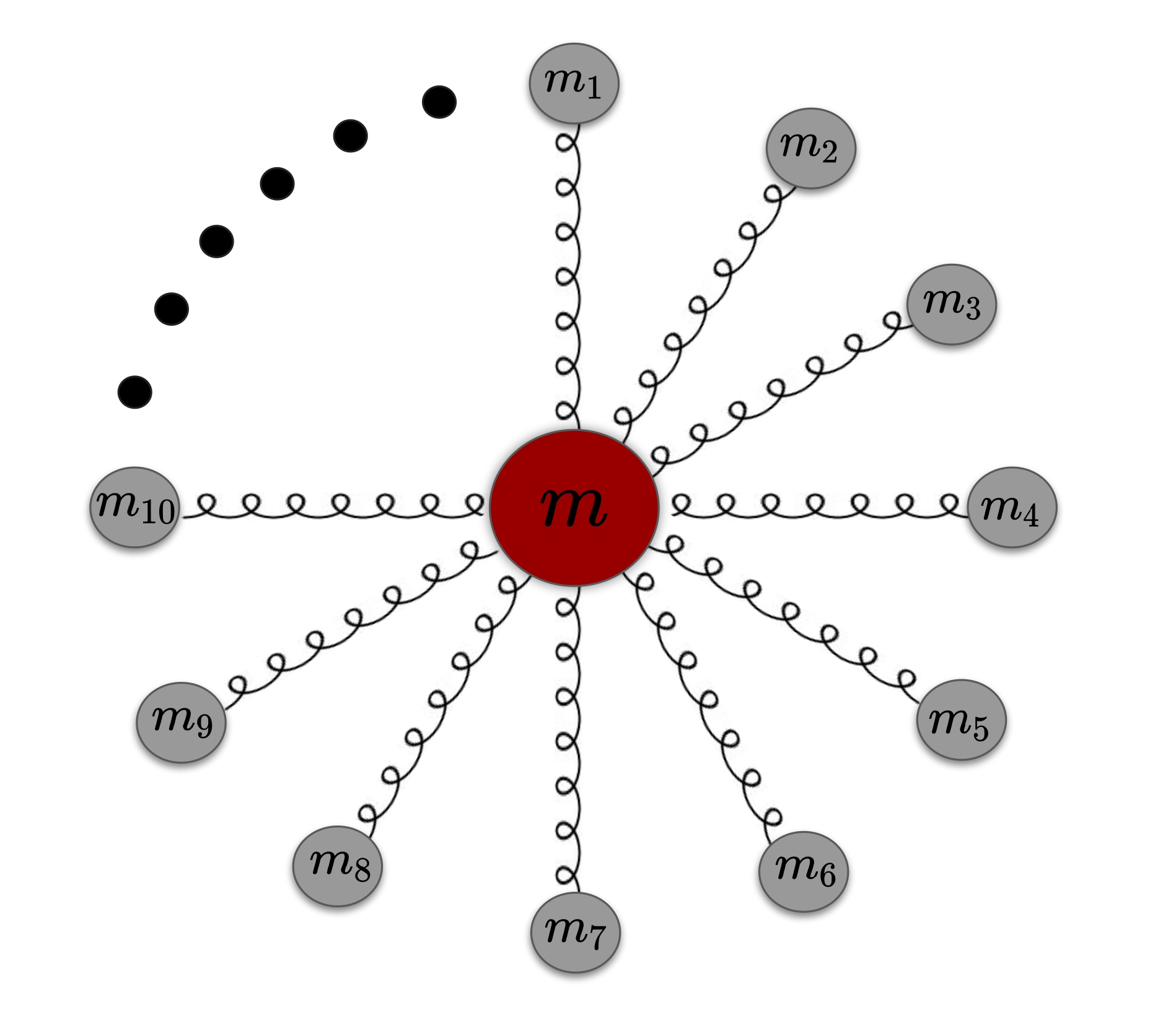}
\caption{The figure shows a large particle of mass \(m\) (the one at the centre) interacting with a set of \(N \gg 1\) independent harmonic
oscillators (represented by the particles at the periphery) with masses $\{m_j\}$ and frequencies $\{\omega_j\};~j=1,2,\dots,N$.}
\label{Mori-Zwanzig}
\end{center}
\end{figure}

Thus, we write the total Hamiltonian as
\begin{equation}
H = \frac{p^2}{2m} + \sum_{j=1}^N \bigg[ \frac{p_j^2}{2m_j} + \frac{ m_j \omega_j^2}{2} \bigg(q_j - \frac{c_j}{m_j \omega_j^2} x \bigg)^2 \bigg],
\end{equation} where we have included an additional contribution \(x^2  \sum_{j=1}^N \frac{c_j^2}{2 m_j \omega_j^2}\) for two reasons. First and foremost, the Hamiltonian is now translationally-invariant, i.e., the system-bath coupling is homogenous in space. Second, the additional contribution precisely cancels a bath-induced `negative frequency shift' on the system (see appendix of \cite{QBM} for the details; this discussion is, however, decades old \cite{mag}). A physical motivation for the
inclusion of the additional contribution is to make the bath excitations look
like displaced harmonic oscillators with the displacement proportional to the system-coordinate itself, yielding the system-bath
coupling, as schematically demonstrated in Fig. (\ref{Mori-Zwanzig}).

\vspace{2mm}

From the Hamilton's equations for the system and bath variables, if one solves those for the bath variables and substitutes back into the Hamilton's equations for the system variables, one gets \cite{QBM}
\begin{eqnarray}
&&\frac{dp}{dt}=-\int_{0}^{t}dt^{\prime}~\frac{p(t^{\prime})}{m}\sum_{j=1}^{N}\frac{c_j^2}{m_j\omega_j^2}\cos[\omega_j(t-t^{\prime})] \\
&&+\sum_{j=1}^{N}c_j \Big(q_j(0)-\frac{c_jx(0)}{m_j\omega_j^2}\Big)\cos(\omega_jt)+\sum_{j=1}^N\frac{c_jp_j(0)}{m_j\omega_j}\sin(\omega_jt) . \nonumber
\end{eqnarray}
The above equation may be rewritten in the form of a generalized Langevin equation (\ref{genlang}) if we identify
 \begin{equation}
 \Theta(t)= \frac{1}{m} \sum_{j=1}^{N}c_j \Bigg[\Big(q_j(0)-\frac{c_jx(0)}{m_j\omega_j^2}\Big)\cos(\omega_jt)+\frac{p_j(0)}{m_j\omega_j}\sin(\omega_jt)\Bigg],
 \end{equation} which depends on the initial coordinates and momenta of the heat-bath oscillators (as well as on the initial particle position). Further, 
  \begin{equation}
 \label{classicalmut}
 \gamma(t) = \frac{\theta(t)}{m}\sum_{j=1}^{N}\frac{c_j^2}{m_j\omega_j^2}\cos(\omega_jt),
 \end{equation}
 where $\theta(t < 0)=0$  and $\theta(t> 0)=1$ is the Heaviside step function. \\
 
 \begin{widetext}
 At this stage, we take the vital step towards introducing irreversibility by considering an ensemble of initial conditions in which the initial coordinate $x(0)$ and the initial momentum $p(0)$ of the particle have given and fixed values for the different members of the ensemble, while the joint distribution of the set $\{q_j(0),p_j(0)\}$ of the initial coordinates and the momenta of the oscillators is given by the canonical-equilibrium distribution (implementing the fact that at the initial instant and also for subsequent times, the bath is in thermal equilibrium): 
 \begin{equation}
\label{dmclass}
\mathcal{P}_0(\lbrace p_j(0)\rbrace,\lbrace q_j(0)\rbrace)=\frac{\exp\left(-\mathcal{H}/k_B T\right)}{\prod_{l=1}^N \int \int dq_l(0) dp_l(0)\exp\left(-\mathcal{H}/k_B T \right)},
\end{equation}
with 
 \begin{equation}
 \mathcal{H}=\sum_{j=1}^N \left[\frac{p^2_j(0)}{2m_j}+\frac{1}{2}m_j\omega_j^2\left(q_j(0)-\frac{c_jx(0)}{m_j\omega_j^2}\right)^2\right].
 \label{HBSB}
 \end{equation}
 \end{widetext}
 Notice that the \(\mathcal{H}\) given above is just the part of the full Hamiltonian that includes the heat bath and the system-bath interactions at time \(t=0\), i.e., \(\mathcal{H} = \big(H_{\rm Bath} - x \sum_{j=1}^N c_j q_j + x^2  \sum_{j=1}^N \frac{c_j^2}{2 m_j \omega_j^2}\big)_{t=0}\). Once this initial preparation is introduced, one finds that the function $\Theta(t)$ is a random function of time whose statistical properties may be deduced by using $\langle q_j (0)\rangle=\langle p_j(0)\rangle=0$, and that $\left\langle \left(q_j(0)-c_j x(0)/(m_j \omega_j^2)\right)^2\right\rangle=k_BT/(m_j\omega_j^2),~~\langle p^2_j(0)\rangle=m_j k_BT$.
Here, the angular brackets denote averaging with respect to the mentioned ensemble, that is, with respect to the distribution~(\ref{dmclass}). We finally obtain 
\begin{eqnarray}
&\langle \Theta(t) \rangle= 0, \\
&\langle \Theta(t)\Theta(t^{\prime}) \rangle= \frac{k_BT}{m} \gamma(t-t^{\prime}). \label{noisecorrclass}
\end{eqnarray}
One may notice that~(\ref{noisecorrclass}) is an explicit expression of the fluctuation-dissipation theorem of the second kind as it directly relates the statistics of fluctuations (the noise) to the dissipation kernel \(\gamma(t)\). In fact, if \(\gamma(t-t') = 2 \gamma \delta(t-t')\), one finds the noise correlation of (\ref{noisestat1}). Such a heat bath is called an Ohmic (or Markovian) bath in which case both the damping and the noise are memoryless. Generally, the kernel \(\gamma(t-t')\) is a function of a memory timescale \(\tau_c\) and falls off such that 
\begin{equation}\label{conditionongamma}
\int_0^\infty  \gamma(\tau) d\tau = \gamma,
\end{equation} which must be finite and positive. One can check that the delta-function (Ohmic) kernel indeed satisfies (\ref{conditionongamma}). Two nontrivial memory kernels prevalent in the literature are
\begin{eqnarray}
\gamma(\tau) &=& \frac{\gamma}{\tau_c} e^{-\tau/\tau_c}, \quad {\rm ~~~~~~~~~(Drude)} \label{Drudekernel} \\
\gamma(\tau) &=& \frac{2\gamma}{\sqrt{\pi}\tau_c} e^{-(\tau/\tau_c)^2}, \quad {\rm (Gaussian)}; \label{Gausskernel}
\end{eqnarray} these go to the Ohmic kernel for \(\tau_c \rightarrow 0\). In what follows, we will discuss the recently-formulated effective-mass approach towards non-Markovian systems where the memory timescale \(\tau_c\) is short but not zero. This approach is distinct from the well-known approach of mapping non-Markovian Langevin equations to Markovian ones by the inclusion of additional variables (see for example, section (1.5) of \cite{zwanbook}). 

\subsection{Effective-mass approach to memory}\label{nmsec3}
Before we proceed further towards discussion of the generalized Langevin equation, we will impose an additional condition given by
\begin{equation}\label{gammacondition2}
\int_0^\infty \tau \gamma(\tau) d\tau = {\rm ~finite},
\end{equation} related to finiteness of the memory time (see \cite{conditions,nonmark3} for more details). For example, for the Drude kernel (\ref{Drudekernel}), one obtains
\begin{equation}
\int_0^\infty \tau \gamma(\tau) d\tau = \gamma \tau_c \int_0^\infty z e^{-z} dz = \gamma \tau_c, \quad \quad z = \tau/\tau_c,
\end{equation} and thus, one finds that the Drude kernel satisfies the requirement (\ref{gammacondition2}). Similarly, for the Gaussian memory kernel, one has
\begin{equation}
\int_0^\infty \tau \gamma(\tau) d\tau = \frac{2 \gamma \tau_c}{\sqrt{\pi}} \int_0^\infty z e^{-z^2} dz = \frac{\gamma \tau_c}{\sqrt{\pi}}, \quad \quad z = \tau/\tau_c. 
\end{equation}
Generally, the memory kernel must fall off faster than \(1/\tau^2\) for large \(\tau\) for both the conditions (\ref{conditionongamma}) and (\ref{gammacondition2}) to hold good. We are now in a position to present the effective-mass formalism \cite{nonmark3,nonmark30}. 

\vspace{2mm}

The basic idea is to perform the following Taylor expansion about \(\tau_c = 0\):
\begin{equation}\label{vtaylor}
v (t - \tau_c u) \approx v(t) - \tau_c u \dot{v}(t), 
\end{equation} where \(u = t/\tau_c\) is a dimensionless parameter and we have neglected terms with higher powers of \(\tau_c\) to a first approximation. Now, the generalized Langevin equation (\ref{genlang}) can be rewritten as
\begin{equation}
\frac{dv}{dt} + \int_0^t \gamma(t') v(t-t') dt' = \Theta (t),
\end{equation} and further, introducing the scaled kernel \(\gamma(t) = \tau_c^{-1} \gamma^*(t/\tau_c)\), we can write
\begin{equation}
\frac{dv}{dt} + \int_0^{t/\tau_c} \gamma^*(u) v(t-\tau_c u) du = \Theta (t),
\end{equation} where \(u = t'/\tau_c\). Substituting (\ref{vtaylor}) and rearranging, we finally arrive at
\begin{equation}\label{langeff1}
\frac{dv}{dt} + \gamma v - \epsilon \tau_c \frac{dv}{dt} = \Theta (t),
\end{equation} where \(\epsilon = \int_0^{t/\tau_c} u \gamma^*(u) du \approx \int_0^{\infty} u \gamma^*(u) du\) as \(\tau_c\) is small; the form of \(\epsilon\) can be derived by taking particular choices for \(\gamma(\tau)\). In arriving at (\ref{langeff1}), we have also used (\ref{conditionongamma}). Thus, we get the Langevin equation
\begin{equation}\label{efflang}
m^* \frac{dv}{dt} + m \gamma v = m \Theta(t), 
\end{equation} where \(m^* = m (1 - \epsilon \tau_c)\) can be interpreted as an `effective' mass with \(m\) being the bare mass. Let us now ask: what is \(\langle \Theta(t) \Theta(t') \rangle\)? It is reasonable to expect that to a first approximation, one should be able to write \(\langle \Theta(t) \Theta(t') \rangle = \Gamma \delta(t-t')\). To find \(\Gamma\), we note that the solution of (\ref{efflang}) turns out to be 
\begin{equation}
v(t) = v_0 e^{-\gamma t/(1 - \epsilon \tau_c)} + \frac{1}{1 - \epsilon \tau_c} \int_0^t e^{-\left(\frac{\gamma}{1 - \epsilon \tau_c}\right)(t-t')} \Theta(t') dt'.
\end{equation}  Thus, the velocity correlation function becomes
\begin{equation}
\lim_{t \rightarrow \infty} \langle v^2(t) \rangle = \frac{(1 - \epsilon \tau_c)\Gamma}{2 \gamma}.
\end{equation}
If we accept that even in the non-Markovian setting, the equilibrium distribution for the velocity is given by a Maxwell distribution, one writes \(\lim_{t \rightarrow \infty} \langle v^2(t) \rangle = k_B T/m\) and this immediately fixes the value of \(\Gamma\) to
\begin{equation}
\langle \Theta(t) \Theta(t') \rangle = \frac{2 \gamma k_B T}{m(1 - \epsilon \tau_c)} \delta(t-t').
\end{equation}Put differently, the above-mentioned noise correlation and the effective-mass Langevin equation (\ref{efflang}), taken together, consistently gives us a Maxwell distribution for the velocity at equilibrium. 

\section{Discussion}\label{Conc}
In this review, we described the many-faceted problem of Brownian motion of a particle suspended in a fluid, primarily relying on the framework of the Langevin equations. After recapitulating the work of Einstein and Langevin which provided the theoretical explanation of this phenomenon, we discussed some recent developments in statistical physics by analyzing this seemingly-simple situation of the dynamics of a tagged particle in a viscous medium. Among what we discussed was stochastic thermodynamics, fluctuation theorems, and non-Markovian Brownian dynamics. Since some of these recent developments have found their place only in specialized monographs but not in textbooks that are accessible to students and researchers who are not active workers in statistical physics, it is our strong belief that the present exposition, although brief, shall provide a simplistic introduction to some of the modern developments in the field of statistical physics to readers with minimal expertise in the field. This belief rests on the fact that our analysis hinges on the simple problem of Brownian motion of a suspended particle in a fluid, a problem that is often familiar to even high-school students. 

\vspace{2mm}

Robert Brown was born in Scotland in 1773, so the year of 2023
was marked as his 250th year of birth. By profession he was a medical
practitioner and a paleobotanist who made significant contributions
to botany largely through the pioneering use of microscopes. It is
through self-designed microscopes that he made the path-breaking
observation of what is called Brownian motion that has left a rich
legacy even after nearly 200 years of his first paper on the subject.
Indeed, Brownian motion is a truly-interdisciplinary topic with
applications to almost all branches of science. Although we restricted
our survey in this review article to the classical aspects of Brownian
motion, a generalization of the underlying ideas to the quantum
domain has led to a huge impact on contemporary interest in open quantum systems. The latter deal with three d's:
Diffusion, Dissipation and Decoherence. Because there is much
attention given today to quantum information leading to computing,
the challenge before the scientists is to minimize the effect of
decoherence. Quantum information is best preserved for nano
systems in which one can maintain coherence phenomena, essential
for computing. However, the flip side is that the nano systems are
invariably under non-negligible dissipative effects due to the
surrounding degrees of freedom which have to be also viewed
quantum mechanically. Therefore, a proper treatment of the
environment requires the use of quantum Langevin equations which
are the quantum versions of what we have reviewed here. One
obviously needs low temperatures for the survival of coherence, but
the flip side is the occurrence of an enhanced quantal timescale
\(\hbar/k_BT\) -- hard to surpass -- and necessarily therefore requiring the
consideration of non-Markovian/memory effects -- not elaborately
discussed as they are beyond the realm of the present study. The
reader can however get a glimpse of the underlying theory from the
monograph \cite{SDG_diff} cited here.\\

\textbf{Acknowledgements:} We thank our recent collaborators -- Jasleen Kaur, Malay Bandyopadhyay, Shamik Gupta, and Subhash Chaturvedi for discussions. S.D. thanks the Indian National Science Academy for support through their
Honorary Scientist Scheme. A.G. thanks the Ministry of Education (MoE), Government of India for support in the form of a Prime Minister's Research Fellowship (ID: 1200454) and also expresses his gratitude to Gert-Ludwig Ingold and Peter Talkner for helpful discussions. A.G. thanks the Department of Physics, Banaras Hindu University for hospitality during the final stages of preparing the manuscript. We are grateful to Jasleen Kaur for carefully reading the manuscript.

\end{document}